\documentstyle[12pt]{article}
\begin{document}
\title{ Black hole entropy:\\ departures from area law}
\author{P. Mitra\thanks{e-mail mitra@saha.ernet.in}\\
Saha Institute of Nuclear Physics\\
Block AF, Bidhannagar\\
Calcutta 700 064, INDIA}
\date{gr-qc/9503042}
\maketitle
\begin{abstract}
The thermodynamic and euclidean functional integral approaches to
black  hole  entropy are discussed. The existence of some freedom
in  the  definition  of  the  entropy  is  pointed  out  and  the
possibility  of  a  departure  from  the semiclassical expression
discussed in the light of quantum corrections. The  semiclassical
area  dependence  of the entropy of matter in the background of a
black hole is also reviewed and shown to break down in  the  case
of  extremal  black  holes.  The cutoff dependence is shown to be
different for the extreme  dilatonic  and  Reissner  -  Nordstrom
black holes.

\end{abstract}
\vfill
\centerline{ Invited Talk delivered at Workshop on}
\centerline{\bf Physics at the Planck Scale, Puri,
December 1994}\vfill
\newpage
\section{Introduction}

It is now widely known that the area of the horizon of a black
hole  can  be  interpreted as an entropy \cite{Bek} and satisfies
all  the  thermodynamical laws. This is not completely understood
in terms of the usual formulation of entropy as a measure of
the number of states available, but the na\"{\i}ve  Lagrangian  path
integral does lead to  a  partition  function  from  which the area
formula for entropy  can  be  obtained  \cite{GH}  by  neglecting
quantum fluctuations.

That formula is supposed to describe the gravitational entropy
corresponding to a black hole.
There have also been investigations on the entropy of quantum
fields  in  black  hole  backgrounds \cite{'tHooft,Uglum}.
The values thus obtained may be considered to be
additional contributions   to  the
entropy  of  the black hole - field system, but the gravitational
entropy itself has occasionally been envisaged to arise  in  this
way. The calculations produce
divergences, with the area of the horizon  appearing as a factor.
This has been interpreted to mean that the gravitational
constant gets renormalized in the presence of the quantum fields \cite{Uglum}.

Of great topical interest is the case of extremal black holes, which
possess peculiarities not present in the corresponding
nonextremal cases \cite{Pres,GK,HHR,Teit}.
For extremal dilatonic black holes,
the temperature is nonzero, but the area vanishes.
For extremal Reissner - Nordstrom black holes,
the temperature is zero, but
the {\it area} is nonzero.
Topological arguments have been presented in the context of Euclidean quantum
gravity to suggest that these extremal  black holes
have zero gravitational entropy in spite of
the nonvanishing area and,
what is more surprising, {\it no definite temperature.}
What has actually been shown, however, \cite{GK,HHR,Teit} concerns
the classical {\it action} of
extremal Euclidean Reissner - Nordstrom
black hole configurations, and it has been
argued to lead semiclassically to a vanishing {\it entropy}.

In this talk
I shall critically reexamine (a) the connection
between the action and the entropy and
(b) the appearance of the area in the expression for the entropy of a scalar
field    in    the   background   of   an  extremal  black  hole.
Both non-extremal and extremal black holes will be considered. New
expressions for entropy will be given in several cases
on the basis of our work \cite{GM}.

To set the stage, let me  start  by  reminding  you  of  the  two
extremal black hole solutions mentioned above.
The metric of the Reissner - Nordstrom spacetime is given by
\begin{equation}
ds^2=-(1-{2M\over r}+{Q^2\over r^2})dt^2+ (1-{2M\over r}+{Q^2\over r^2})
^{-1}dr^2 +r^2d\Omega^2
\end{equation}
in general, with $M$ and $Q$ denoting the mass and the charge
respectively.  This  is  a  solution  of  the  Einstein - Maxwell
equations. There are apparent singularities at
\begin{equation}
r_\pm=M\pm\sqrt{M^2-Q^2}
\end{equation}
provided $M\ge Q$. Cosmic censorship dictates that this
inequality holds and then there is a horizon at $r_+$.
The limiting case when $Q=M$ and $r_+=r_-$ is referred to as
the extremal case.

There is also an extremal case of the dilatonic black hole.
The usual four - dimensional model \cite{dil} is
\begin{equation}
S={1\over 16\pi}\int d^4x\sqrt{- g}( R-2 g^{\mu\nu}\nabla_{\mu}\phi
\nabla_{\nu}\phi)\label{2},
\end{equation}
where  $\phi$  is the massless dilaton field, $R$  the
scalar  curvature   and   $g_{\mu\nu}$   the  metric.
Electromagnetic  interactions are brought in by including the term
\begin{equation}
-{1\over 32\pi}\int d^4x\sqrt{- g}e^{-2\phi} g^{\mu\lambda}
g^{\nu\rho}F_{\mu\nu}F_{\lambda\rho}
\end{equation}
in the action. Exact black hole
solutions of this model have been found with non-zero magnetic charge
and  angular momentum.

The  black hole solution with zero angular momentum strongly resembles
the Schwarzschild solution of standard general relativity.
\begin{eqnarray}
ds^2&=& g_{\mu\nu}dx^{\mu}dx^{\nu}\nonumber\\
&=&-(1-{2M\over r})dt^2 +(1-{2M\over r})^{-1}dr^2 +r(r-a)d\Omega^2
\nonumber\\ e^{-2\phi}&=&e^{-2\phi_0}(1-{a\over r})\nonumber\\
F_{\theta\varphi}&=&Q\sin\theta
\label{dbh}\end{eqnarray}
where  $M$ is the mass of the black hole, $Q$ its magnetic charge,
the parameter $a$ is defined by
\begin{equation}
a={Q
^2\over 2M}e^{-2\phi_0}\end{equation}
and $\phi_0$ is an arbitrary constant.
This black hole
has as usual a horizon at $r=2M$. An interesting feature
is that a curvature singularity occurs
at   $r=a$.  The  so-called  extremal solution corresponds to the
coincidence of these two regions and thus has $a=2M$. This
extremal limit is interesting from the point of view of entropy because the
area $4\pi 2M(2M-a)$ of the horizon vanishes.

As mentioned above,
two different contributions to entropy will be discussed
in this talk, and their relation can be understood as
follows. As argued in \cite{Hawk} the partition function for the
system can be defined by a Euclidean path integral for the
gravitational action coupled with matter fields. The dominant
contribution will come from the classical solutions of the action.
So one expands the Euclidean action,
which involves the metric,  additional fields (the dilaton
and/or the electromagnetic field) and an external scalar matter field, as
\begin{equation}
S_E[g,\Phi,\varphi]=S_E[g_0,\Phi_0]+S_2[\delta g,\delta\Phi,\delta\varphi;
g_0,\Phi_0]+\cdots,
\end{equation}
where
\begin{equation}
g_{\mu\nu}=g_{0\mu\nu}+\delta g_{\mu\nu},\qquad\Phi=\Phi_0+
\delta\Phi,\qquad\varphi=\varphi_0+\delta\varphi.\end{equation}
Here the subscript 0 denotes the classical value and
$\delta $ denotes the fluctuations.  $S_2$ is the part
quadratic in the fluctuations.
The background matter field is taken as $\varphi_0=0$ {\it i.e.}
the contributions of these fields are subtracted out while
calculating the quantum corrections due to the matter fields.
$S_2$ can be decoupled into the
gravitational ($\delta g, \delta\Phi$) and the matter
($\delta\varphi$) sectors. The entropy arising from
the partition function defined by ignoring all fluctuations
is the semiclassical entropy. Consideration
of the gravitational part of $S_2$ leads to quantum
corrections to the gravitational entropy.
Inclusion of the matter part of $S_2$ adds
the entropy of the scalar field in the background of the black hole.

\section{Gravitational entropy}
For the  simplest  black  hole,  namely  the  one  discovered  by
Schwarzschild, the Hawking temperature is given by
\begin{equation}
T={1\over 8\pi M},\end{equation}
where  $M$  is the mass of the black hole. Accordingly, the first
law of {\it thermodynamics} can be written as
\begin{equation}
dM=TdS=(8\pi M)^{-1}dS,\end{equation}
which shows that the entropy must be $4\pi M^2$ upto an  additive
constant,  {\it  i.e.},  essentially  a  quarter  of the area. No
analogy is involved here, and the  standard  result  is  obtained
directly   from  thermodynamics.  There is no scope for ambiguity
at  this  level  of  approximation   defined   by   the   Hawking
temperature. Unfortunately this is no longer the case if we go on
to  more complicated black holes. We shall demonstrate below that
thermodynamics allows some freedom  in  the  expression  for  the
entropy  in the case of black holes depending on extra parameters
like charge or angular momentum. Furthermore, the expression used
for the Hawking temperature is a semiclassical one and  therefore
subject  to  higher  quantum corrections. This can be expected to
lead to quantum corrections to the entropy. We shall indicate how
all these extra terms can be obtained, at least in principle,  by
calculating some functional integrals.

A generalized Smarr   formula   \cite{Smarr,BCH}   can
be written down for charged black holes as
\begin{equation}
M={TA\over 2}+\Phi Q.\label{Smarr}\end{equation}
Here $\,A\,$ is the area of the horizon and $\,\Phi\,$ is the
analogue of the
chemical potential for the electric or magnetic charge given by
\begin{equation}
\Phi=\left( {\partial M\over\partial Q}\right)_A
{.}\end{equation}
The formula (\ref{Smarr}) is the integrated form of the
first law of black hole physics \cite{BCH}
\begin{equation}
Td(A/4)=dM-\Phi dQ.\label{a}\end{equation}
By comparing this with the first law of thermodynamics
\begin{equation}
TdS=dM-\Phi dQ\end{equation}
one may be tempted to conclude that the entropy is  simply  $\,A/4\,$
upto an additive constant \cite{Pres}.

It  has  been  tacitly assumed that the $\,\Phi\,$-s entering the
above two equations are  the  same.  Whereas  the  definition  of
$\,\Phi\,$  given  above  is applicable to the first law of black
hole physics, the first law of thermodynamics  should  really  be
written as
\begin{equation}
\tilde TdS=dM-\tilde\Phi dQ\label{s}\end{equation}
where $\,\tilde\Phi\,$ is
\begin{equation}
\tilde\Phi=\left( {\partial M\over\partial Q}\right)_S.\end{equation}
In  assuming  $\,\tilde\Phi\,$  to  be the same as $\,\Phi\,$ one in
effect  assumes  that  differentiation at constant entropy is the
same  as  differentiation  at  constant  area,  which  makes  the
derivation of the area formula for the entropy somewhat circular.
We use different symbols to allow for a possible difference.
One  should  also  allow  for  higher  quantum corrections to the
temperature and therefore write $\,\tilde T\,$.

Subtracting (\ref{a}) from (\ref{s}) we find
\begin{equation}
d(S-{A\over 4})=\beta_qdM-({\tilde\Phi\over\tilde T}-
{\Phi\over T})dQ,\end{equation}
where
\begin{equation}
\beta_q=
{1\over\tilde T}-{1\over T}.\end{equation}
We set
\begin{equation}
S={A\over 4}-F(Q, M),\end{equation}
where the function $F$ is undetermined except for the requirement that
\begin{equation}
{\partial F\over\partial M}=-\beta_q.\label{nu}\end{equation}
Furthermore,
\begin{equation}
{\partial F\over\partial Q}=
{\tilde\Phi\over\tilde T}-{\Phi\over T}.\end{equation}
But this equation is not a restriction on $F$ because $\tilde\Phi$ is not
known. It can be used to fix the latter quantity if $F$ can be found.

$F$ can be sought  to  be  determined  from  the
partition function
\begin{equation}
Z=e^{-I_E}\end{equation}
where $\,I_E\,$ is an effective action defined from the Euclidean
functional integral \cite{GH}. $\,Z\,$ can also be interpreted as
the grand canonical partition function
\begin{equation}
Z=e^{-W/\tilde T},\end{equation}
where
\begin{equation}
W=M-\tilde TS-\tilde\Phi Q.\end{equation}
In the stationary  phase approximation the functional integral is
taken  to  be  dominated  by  the  classical  configuration   and
$\,I_E\,$ is approximately equal to $\,A/4\,$ \cite{Pres,Ort}.
So the full effective action including fluctuations can be written as
\begin{eqnarray}
I_E&=&{A\over 4}+I_{\rm corr}\nonumber\\
   &=&{1\over 2T}(M-\Phi Q)
+I_{\rm corr}.
\end{eqnarray}
{}From these equations we can see that
\begin{equation}
{M-\tilde TS-\tilde\Phi Q\over\tilde T}=
{1\over 2T}(M-\Phi Q)+I_{\rm corr}\end{equation}
By  substituting the expressions for $\,S\,$ and $\,\tilde\Phi\,$ we find
\begin{equation}
M(\beta_q+{1\over 2T})-{A\over 4}-{\Phi Q\over 2T}=
I_{\rm corr}+Q{\partial F\over\partial Q}-F.\end{equation}
Comparison with the Smarr formula (\ref{Smarr}) and (\ref{nu})
yields the result
\begin{equation}
F-Q{\partial F\over\partial Q}-M{\partial F\over \partial M}
=I_{\rm corr}.\label{F}\end{equation}
In the semiclassical approximation $\,I_{\rm corr}\approx 0.$
Then  $\,F\,$  need not vanish, but {\it can  be}  zero, yielding the usual
expression for the entropy. Generally, if $\,I_{\rm corr}\,$ is
calculated from the Euclidean functional integral, (\ref{F})  can
be  formally solved as
\begin{equation}
F=(1-Q{\partial  \over\partial  Q}-M{\partial  \over
\partial  M})^{-1}I_{\rm  corr}
+F_0,\end{equation}
where $F_0$ is a solution of the homogeneous equation. It cannot be an
arbitrary solution of the homogeneous equation because
of the restriction imposed by (\ref{nu}). $F_0$ has to be
independent of $M$, and to satisfy the homogeneous equation
it has to be a linear function of $Q$. This piece of $F$
exposes  the  freedom  in  the definition  of  the  entropy, whereas
the first piece provides quantum  corrections to both the  entropy
and  the temperature.
If $F$ is to be proportional to the area, which is a
homogeneous function of $M$ and $Q$, it is clear
that $I_{\rm  corr}$ itself must be proportional to the area.

This general discussion  about charged  black  holes
relates to the non-extremal case. The extremal
case is special because of
the newly discovered properties of extremal black
holes \cite{GK,HHR,Teit} which cannot be continuously deformed
into nonextremal black holes and which, moreover, can be in equilibrium
at arbitary temperatures. We shall now investigate the possibility of
an  ambiguity  in the definition of the semiclassical entropy for
such black holes, {\it i.e.}, the analogue  of  the  $F_0$  noted
above.

The first law of thermodynamics is (\ref{s}),
as before.
In  this  extremal  case,   $Q=\alpha  M$,  where  $\alpha$  is a
constant  depending  on  the type of the black hole (dilatonic or
Reissner - Nordstrom). Then
\begin{equation}
\tilde TdS=(1-\alpha\tilde\Phi)dM.\label{FL}
\end{equation}

We now appeal to the vanishing Euclidean action \cite{Teit}.
The partition  function can be approximately written as
\begin{equation}
Z=e^{-I},
\end{equation}
with $I=0$. Now note that this is the grand canonical partition
function \cite{GH} because the charge
$Q$ is  variable. It is not an {\it independent}
variable if we consider black holes which remain extremal, but it has to
change in processes where the energy $M$ changes. So we write the
usual formula
\begin{equation}
Z=e^{-W/\tilde T},
\end{equation}
with
\begin{equation}
W=M-\tilde TS-\tilde\Phi Q.
\end{equation}
Noting that $W$ has to vanish, we arrive at the relation
\begin{equation}
\tilde TS=(1-\alpha\tilde\Phi)M.
\end{equation}
Comparing with (\ref{FL}), and assuming the factors to
be nonvanishing, we conclude that
\begin{equation}
{dS\over S}={dM\over M},
\end{equation}
{\it i.e.,}
\begin{equation}
S=kM, \label{k}
\end{equation}
where $k$ is an undetermined constant. If $k$ vanishes,
we get a vanishing entropy \cite{HHR,Teit},
but (\ref{k}) is more generally valid.
It is crucial here to abandon the na\"{\i}ve result $\alpha\Phi=1$
which might be obtained by assuming continuity  of   non - extremal
and extremal black holes. As has been argued in
\cite{HHR}, the temperature cannot be fixed.
The potential too cannot be fixed, but it is related
to the temperature by the formula
\begin{equation}
\alpha\tilde\Phi=1-k\tilde T.
\end{equation}
The only definite result is that the thermodynamical
entropy, in contradistinction to the action, is
given by the mass of the extremal black hole upto a
constant which may, however, be zero. Note that this  is  at  the
semiclassical level itself.

Perhaps it is more interesting to look at
quantum corrections, but these are not reliably known in 3+1 dimensional cases.
We consider therefore a 2+1 dimensional black hole \cite{BTZ}. The
rotating version is described by the metric
\begin{equation}
ds^2=-f^2dt^2 +f^{-2}dr^2+r^2(d\phi-{J\over 2r^2}dt)^2,\end{equation}
where
\begin{equation}
f^2=-8GM+{r^2\over l^2}+{16G^2J^2\over r^2},\end{equation}
and $l^{-2}$ is a cosmological constant. The constant $G$ is retained
here in contrast to the rest of the talk because of the unusual units used
in \cite{BTZ}.
The outer horizon is at $r_+$, where
\begin{equation}
r_+^2=4GMl^2\left[1+\sqrt{1-{J^2\over M^2l^2}}\right].\end{equation}
The Hawking temperature is
\begin{equation}
T={r_+^2-4GMl^2\over \pi r_+l^2}.\end{equation}
The first law of black hole physics can be written in the
form
\begin{equation}
Td{\pi r_+\over 2G}=dM-\Omega dJ.\label{A2}\end{equation}
The {\it area} is $2\pi r_+$ because of the 1-dimensional nature of
the horizon here. This
equation can be directly checked from the expression for $r_+$.
The analogue of the chemical potential for the angular momentum is
\begin{equation}
\Omega=\left({\partial M\over\partial J}\right)_A={4GJ\over r_+^2}.
\end{equation}
Comparing (\ref{A2}) with the first law of thermodynamics
\begin{equation}
\tilde TdS=dM-\tilde\Omega dJ,\end{equation}
we can write
\begin{equation}
d(S- {\pi r_+\over 2G})=\beta_qdM
-({\tilde\Omega\over\tilde T}-{\Omega\over
T})dJ.\end{equation}
As before,
\begin{equation}
S= {\pi r_+\over 2G}-F(J, M),\end{equation}
with $\,F(J, M)\,$
as yet undetermined except for (\ref{nu}).
The connection between  the  unknown  function  $\,F\,$  and  the
effective action $\,I_E\,$ is given by
\begin{equation}
I_E=F-J{\partial F\over\partial J}-M{\partial F\over\partial M}
-{\pi r_+\over 4G}\end{equation}
in view of the Smarr formula which in this case reads
\begin{equation}
M={\pi r_+T\over 4G}+\Omega J.\end{equation}
If  $F$  is  to  be proportional to $r_+$, which is a homogeneous
function of $M$ and $J$, it is clear that $I_E$  itself  must  be
proportional to $r_+$. Indeed, such an  expression  for  a
quantum  correction  has  been  given  in \cite{BTZ}:
\begin{equation}
I_E\approx  -{\pi  r_+\over 2G}-{2\pi  r_+\over  l} +{M\over T} -{4GJ^2\over
r_+^2 T}.\end{equation}
It follows that $F$ must satisfy the equation
\begin{equation}
F-J{\partial F\over\partial J}-M{\partial F\over\partial M}=
-{2\pi r_+\over l}.\end{equation}
This equation can be formally solved as
\begin{eqnarray}
F&=&(J{\partial \over\partial J}+M{\partial
\over\partial M}-1)^{-1}{2\pi r_+\over l}
\nonumber\\&=&-{4\pi r_+\over l} +F_0.\end{eqnarray}
Accordingly, a possible answer for the entropy is
\begin{equation}
S={\pi r_+\over 2G}+{4\pi r_+\over l}-F_0,\end{equation}
which is similar to, though not identical with, the expression
suggested in \cite{BTZ} by comparison  with  the  quantum  action.
They in effect ignored not only the possibility of the solution $F_0$
of  the  homogeneous  equation,  but  also  the  presence  of the
derivatives of $F$ in the equation.

If $\hbar$ is reinstated and $F_0$ ignored, we can write
\begin{equation}
S={2\pi r_+\over 4\hbar G}+{4\pi r_+\over l}.\end{equation}
Using (\ref{nu}), one can also find the corrected temperature:
\begin{equation}
\tilde   T^{-1}={\pi   r_+l^2\over r_+^2-4GMl^2}({1\over    8\hbar
G}+{1\over l}).\end{equation}

To  sum  up,  we  have  demonstrated that only the simplest black holes  have
their  entropies  uniquely  determined  by  the  first   law   of
thermodynamics at the semiclassical level of approximation.
More generally, an expression for the entropy can be written down in terms
of a function of the mass, charge, angular
momentum and so on. Differential equations relate these functions
to the Euclidean functional integrals for  the  black  holes, but
there  is  no  unique  solution.  However, if  one  knows  higher  quantum
corrections to these functional integrals,
the   equations   may  be  used  to  determine the corresponding
corrections to both the entropy and the temperature.

\section{Matter outside black hole}
To  study the entropy of a scalar field in the background of
a black  hole   we   employ   the
brick-wall boundary condition \cite{'tHooft}. In this model the
wave function is cut off just outside the horizon. Mathematically,
\begin{equation}
\varphi(x)=0\qquad {\rm at}\;r=r_h+\epsilon
\end{equation}
where   $\epsilon$  is a small, positive, quantity and signifies
an ultraviolet cut-off. There is also an infrared cut-off
\begin{equation}
\varphi(x)=0\qquad {\rm at}\;r=L
\end{equation}
with   $L>>r_h$.

We consider a static, spherically symmetric black hole spacetime
with the metric
\begin{equation}
ds^2=g_{tt}(r)dt^2 +g_{rr}(r)dr^2+ g_{\theta\theta}(r)d\Omega^2,
\end{equation}
and study spinless particles bounded by the brick wall at $r= r_h
+\epsilon$ and the long distance cutoff at $r=L$.
An $r$- dependent  radial  wave  number  can  be  introduced  for
particles  with mass $m$, energy $E$ and orbital angular momentum
$l$ by
\begin{equation}
k_r^2(r,  l,  E)=  g_{rr}[-g^{tt} E^2 -
{l(l+1)g^{\theta\theta}} -m^2].
\end{equation}
Only such values of $E$ are to be considered here that the  above
expression  is  nonnegative. The values are further restricted by
the semiclassical quantization condition
\begin{equation}
n_r\pi=\int_{r_h+\epsilon}^L~dr~k_r(r, l, E),
\end{equation}
where the radial quantum number $n_r$ has to be a nonnegative integer.

The free energy $F$ at inverse temperature $\beta$
is given by a sum over single particle states
\begin{eqnarray}
\beta F&=&\sum_{n_r, l, m_l}\log(1-e^{-\beta E})\nonumber \\
&\approx  &  \int  dl~(2l+1)\int  dn_r\log   (1-e^{-\beta   E})
\nonumber\\
&=&-\int  dl~(2l+1)\int d(\beta E)~(e^{\beta E} -1)^{-1} n_r \nonumber\\
&=& -{\beta\over\pi}\int  dl~(2l+1)
\int dE~(e^{\beta E} -1)^{-1}\int_{r_h+\epsilon}^L
dr~g_{rr}^{1/2}\nonumber\\
&& \sqrt{-g^{tt}E^2-{l(l+1)g^{\theta\theta}}-m^2} \nonumber\\
&=&  -{2\beta\over  3\pi}\int_{r_h+\epsilon}^L  dr~  g_{rr}^{1/
2}g_{\theta\theta}(-g_{tt})^{-3/2}
\nonumber\\&& \int dE~(e^{\beta E} -1)^{-1}
[E^2+g_{tt}m^2]^{3/2}.
\label{long}\end{eqnarray}
Here  the  limits  of  integration  for  $l, E$ are such that the
arguments  of  the  square  roots  are   nonnegative.   The   $l$
integration  is  straightforward  and has been explicitly carried
out. The $E$ integral can be evaluated only approximately.

Because  of  the  asymptotically flat nature of the metric of the
spacetime containing the black hole,
the contribution to the $r$ integral from  large  values  of  $r$
corresponds to the  expression  for  the  free  energy  valid  in  flat
spacetime:
\begin{equation}
F_0=-{2\over 9\pi}L^3\int_m^\infty dE{(E^2-m^2)^{3/2} \over
e^{\beta E} -1}.
\end{equation}
This piece is not relevant.
The contribution of the black hole  is  singular  in  the  limit
$\epsilon\to  0$.  The  leading singularity is obtained by taking
the metric coefficients, multiplied by appropriate powers of $(r-
r_h)$  to  make them finite at the horizon, outside the integral.
For a non-extremal black hole, $g_{rr}$ has a linear  singularity
and  $g_{tt}$  a linear zero at $r=r_h$, while $g_{\theta\theta}$
is regular there. Thus,
\begin{equation}
F_{sing}\approx -{2\pi^3\over  45\epsilon\beta^4}
[(r-r_h)g_{rr}]^{1/2}(-{g_{tt}\over
r-r_h})^{-3/2}g_{\theta\theta}|_{r=r_h},
\end{equation}
where  the lower limit of the $E$ integral has been approximately
set equal to zero. If  the  proper  value  is  taken,  there  are
corrections  involving  $m^2\beta^2$  which will be ignored here.

The entropy due to the black hole can be obtained from the formula
\begin{equation}
S=\beta^2 {\partial F\over\partial\beta}.
\end{equation}
This gives
\begin{equation}
S_{sing}=  {8\pi^3\over  45\beta^3
\epsilon}
[(r-r_h)g_{rr}]^{1/2}(-{g_{tt}\over
r-r_h})^{-3/2}g_{\theta\theta}|_{r=r_h}.
\end{equation}
To  put  this  expression  in  a more familiar form, we write the
Hawking temperature as
\begin{eqnarray}
{1\over\beta}&=&{1\over 2\pi}
(g_{rr})^{-1/ 2}{\partial \over\partial r}(-g_{tt})^{1/2}|_{r=r_h}
\nonumber\\ &=&{1\over 4\pi}(g_{rr})^{-1/ 2}(-g_{tt})^{-1/ 2}
{\partial \over\partial r}(-g_{tt})|_{r=r_h}
\nonumber\\ &= &{1\over 4\pi}[(r-r_h)g_{rr}]^{-1/ 2}(-{g_{tt}\over
r-r_h})^{1/2}|_{r=r_h}.
\end{eqnarray}
It is also necessary to measure the width of the brick
wall in terms of the proper radial
variable $\tilde r$ defined by $d\tilde r^2=g_{rr}dr^2$:
\begin{equation}
\tilde\epsilon=\tilde r(r_h+\epsilon)-\tilde r(r_h)
\approx 2\epsilon^{1/2}[(r-r_h)g_{rr}]^{1/2}|_{r=r_h}.
\end{equation}
On making these substitutions, we find
\begin{equation}
S_{sing}=  {1\over  90 \tilde\epsilon^2}
g_{\theta\theta}|_{r=r_h}=  {1\over  360\pi\tilde\epsilon^2}
{\rm Area},
\end{equation}
in agreement with known results \cite{'tHooft}.

The above derivation crucially depends on the  behaviour  of  the
metric  coefficients  near  the  horizon  and  is  valid only for
non-extremal black holes. For extremal cases,  the  behaviour  is
different and the area formula does not emerge.

For the extremal dilatonic black hole,
$F$  has  a  logarithmic  singularity.  This  is  present  in the
non-extremal case  as well, but it is in general
ignored because of the presence of the linearly  divergent  term.
However,  the  linear term vanishes when $a=2M$, {\it
i.e.}, when the black hole becomes extremal. In  this  case,  the
logarithmic  term  is  the  dominant  one.  It   arises   because
$g_{\theta\theta}$ vanishes linearly at the horizon and has to be
kept inside the $r$ integral in the last line of (\ref{long}). One obtains
\begin{equation}
F_{dil}\approx -{\pi^3\over  45M}\log({2M\over\epsilon})
({2M\over\beta})^4
\end{equation}
in the same approximation as above. Correspondingly,
\begin{equation}
S_{dil}=  {1\over  360} \log {(4M)^2
\over \tilde\epsilon^2}.
\end{equation}
As the area of the horizon  vanishes,  one  might  have
expected  the  entropy  to vanish altogether. What does happen is
that  the  linear  divergence  vanishes,  but   the   logarithmic
divergence,  which  is  of  course  weaker,  stays  on. A similar
logarithmic divergence  is  known  to  occur  if  the  theory  is
truncated to (1+1) dimensions \cite{Uglum}. Our calculation shows
that this is already present in (3+1) dimensions.

For an extremal Reissner - Nordstrom black hole, the  singularity
of  the $r$ integral in the last line of (\ref{long}) becomes stronger
because $g_{rr}$ has  now a  quadratic  singularity  and  $g_{tt}$  a
quadratic zero at the horizon. One obtains
\begin{equation}
F_{RN}\approx -{2\pi^3r_+^2\over  135\epsilon^3}
({r_+\over\beta})^4.
\end{equation}
The contribution to the entropy due to the
presence of the black hole is
\begin{equation}
S_{RN}=  {8\pi^3\over  135}({r_+\over\beta})^3({r_+
\over\epsilon})^3.\label{S}
\end{equation}
The formula for the Hawking temperature of the Reissner -
Nordstrom black hole is
\begin{equation}
T={r_+-r_-\over 4\pi r_+^2}.
\end{equation}
This expression vanishes in the extremal case where $r_+ =r_-$.
If the vanishing temperature is inserted,
the expression (\ref{S}) for the entropy also
vanishes and this has been the understanding about this
entropy until  recently. However, the temperature may
be nonvanishing, because of
the new observation \cite{HHR} that the Euclidean
solution can be identified with an {\it arbitrary} period $\beta$.
For a general $\beta$ (\ref{S}) is nonzero and nominally cubically
divergent, whereas nonextremal black holes have only a linear
divergence in terms of the   cutoff  $\epsilon$.  In
this extremal case, the cutoff in the proper radial
variable $\tilde r$ defined as above goes like -$\log~\epsilon$,
so that the true divergence is exponential, {\it i.e.},
much stronger than in the non-extremal case.

This is a rather surprising result.
Whereas the previous section
suggests that the gravitational entropy of an extremal black hole
of the Reissner - Nordstrom type is essentially given by the
mass, the entropy of the scalar field in this background is  even
{\it more} singular than in the nonextremal case. The background
of an extreme dilatonic black hole also gives a nonzero result
when a zero might have been expected,
but there the leading linear singularity does drop out and only a
milder, logarithmic term remains.

\section*{Acknowledgments}
This talk is based on work done in collaboration with Amit Ghosh.

\end{document}